\documentclass[journal]{IEEEtran}
\usepackage{amsmath,amssymb}
\usepackage{graphicx}
\usepackage{balance}
\usepackage{cite}
\usepackage[justification=centering]{caption}

\begin{document}

\title{Spatial Upper Bound of Radiated Power in Active Antenna Systems}
\author{
\IEEEauthorblockN{NUSSBAUM Dominique$^{(1)}$, RIZK Christ$^{(1)}$,  SEGUENOT Eric$^{(1)}$,  KALTENBERGER Florian$^{ (2)}$, \\ MORO Andrea$^{ (3)}$, SINICCO Alessandro$^{(3)}$ and POMETCU Laura$^{ (4)}$\\}
\IEEEauthorblockA{
    $^{(1)}$INNOV/ANT Department, ORANGE, Sophia Antipolis, France\\
    $^{(2)}$Communication systems Department, Eurecom, Sophia Antipolis, France\\
     $^{(3)}$Andrew, Agrate Brianza, Italy\\
     $^{(4)}$Agence Nationale des FRequences (ANFR), France}
}
\date{February 2026}
\maketitle

\begin{abstract}
The assessment of unwanted radiated emissions from Active Antenna Systems (AAS) has become a critical issue in adjacent-band coexistence scenarios.
In this paper, we establish the existence of a deterministic spatial upper bound on the radiated
power of active antenna arrays. We show that the maximum radiated power always occurs in the boresight direction, irrespective of frequency or signal nature (useful signal, nonlinear distortion, or noise), or instantaneous beamforming configuration, thereby defining a conservative spatial upper bound whose angular envelope is solely determined by the elementary radiating building block of the antenna architecture, i.e., the element or sub-array radiation pattern.

Starting from a two-element array with third-order nonlinearities, we derive the spatial envelope
and extend the result to realistic AAS architectures.
The theoretical findings are validated by over-the-air (OTA) measurements performed on a 3.5 GHz
Massive Multiple-Input Multiple-Output (MIMO) antenna. The proposed approach offers a simple, robust, and
measurement-oriented methodology for coexistence assessments involving beamformed radio
systems.
\end{abstract}

\section{Introduction}
\label{sec:introduction}
5G Massive MIMO base stations employ Active Antenna Systems (AAS) with digital beamforming to 
achieve high spectral efficiency. While enabling sharp in-band 
directivity, these architectures raise critical questions regarding 
unwanted radiated emissions in adjacent-band coexistence scenarios, 
particularly with safety-critical systems like aeronautical radio 
altimeters. In such contexts, conventional interference assessment approaches,
initially devised for passive or weakly directive antennas, may become overly complex or insufficiently
conservative when applied to dynamically beamformed AAS.

\subsection{Related work and limitations}
Recent work \cite{Larsson2018OOBClarified} showed that nonlinear distortion of power 
amplifiers (PA) can remain correlated across antenna branches and be 
beamformed in the same direction as the intended signal. Our 
measurements \cite{Rizk2026SpatialUpperBound} on a 32T32R Massive MIMO radio unit confirmed 
that spatial directivity of Out-Of-Band (OOB) emissions depends 
on PA operating regime and Digital Predistortion (DPD) effectiveness.

However, a complementary regulatory question remains: can one 
establish a simple, conservative bound on maximum radiated emission 
in any direction, regardless of beamforming configuration, frequency, 
or signal nature?

\subsection{Contribution and key idea}
This paper addresses the above question by establishing that maximum radiated emission always 
occurs at boresight, defining a deterministic spatial upper bound 
whose angular envelope is solely determined by the element or 
sub-array radiation pattern, independent of frequency and signal 
nature. Starting from a two-element array, we extend the result 
to realistic AAS and validate it via OTA measurements on a 3.5 GHz 
Massive MIMO antenna.

\subsection{Methodology and organization}
 We start from elementary array configurations to provide an intuitive
derivation of the spatial envelope, and then extend the reasoning to realistic Massive MIMO AAS
architectures.

The remainder of this paper is organized as follows.
Section II introduces the two-element reference model. Section III derives the spatial upper bound. Section IV deals with the multi-user transmission, section V extends the result to realistic AAS architectures.
Section VI discusses measurement validation and implications. Section VII concludes the paper.

\section{Two-Element Array and Third-Order Nonlinearity}
\label{sec:two_element_array}
\subsection{Two-element array model}

We consider the simplest active antenna configuration composed of two identical radiating elements.
\begin{figure}[h] 
    \centering
    \includegraphics[width=0.7\columnwidth]{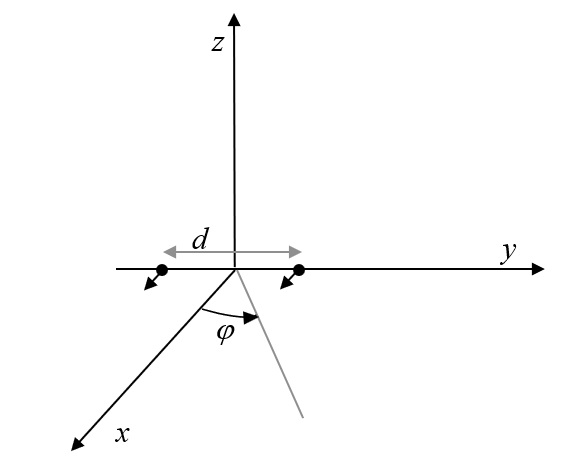} 
    \caption{two-element array positioned along the y-axis.}
    \label{model}
\end{figure}

The elements are aligned along the y-axis and separated by a 
distance $d=\lambda/2$, where $\lambda$ denotes the wavelength 
at the carrier frequency. This half-wavelength spacing is 
standard in practical antenna arrays as it provides a good 
compromise between spatial sampling (avoiding grating lobes) 
and physical compactness. Mutual coupling is neglected, which 
is justified for $\lambda/2$ spacing in well-designed arrays 
with low cross-coupling between elements (typically $< -15$ dB 
in practical Massive MIMO implementations \cite{balanis2016}). 
Ideal amplitude and phase control is assumed, consistent with 
modern digital beamforming architectures where calibration 
procedures ensure accurate control of excitation coefficients. 
The analysis is restricted to the horizontal plane defined by 
$\theta = 90^\circ$, which is the most relevant configuration 
for terrestrial coexistence scenarios involving ground-based 
systems such as 5G base stations and radio altimeters operating 
in adjacent frequency bands. The radiation characteristics are 
studied as a function of the azimuth angle $\varphi \in [0,2\pi]$.
The radiation characteristics of the elementary radiator follow the 3GPP antenna modeling
framework
\cite{3GPP38803}.
The gain in the plane $\theta = 90^\circ$ is given by:
\begin{equation}
A_E(\varphi) = G_{E,max} +A(\varphi),
\tag{1}
\end{equation}
where $G_{E,\max}$ is the element peak gain and:

\begin{equation}
A(\varphi)
=
-
\min\!\left\{
12\left(\frac{\varphi}{\varphi_{3\mathrm{dB}}}\right)^2,
A_m
\right\},
\tag{2}
\end{equation}
where $\mathrm{A}_m$ is the front to back ratio and  $\varphi_{3\mathrm{dB}}$ is the horizontal half power beamwidth.
For a coherent frequency component
\cite{balanis2016}, the radiation pattern of the array in the horizontal plane is expressed as:
\begin{equation}
G(\varphi,\Delta\Phi)
= A_E(\varphi)
+ 20\log_{10}|AF(\varphi,\Delta\Phi)|,
\tag{3}
\end{equation}
where $\Delta\Phi$ denotes the relative excitation phase between the two radiating elements and the array factor in the horizontal plane can be written as:
 
\begin{equation}
AF(\varphi,\Delta\Phi)
= 1 + e^{j\left(\psi(\varphi)+\Delta\Phi\right)},
\tag{4}
\end{equation}

where the geometrical phase term $\psi(\varphi)$ accounts for the propagation path difference
between the two elements in the observation direction $(\varphi)$. For a two-element array
with inter-element spacing $d$ and wave number $k = 2\pi/\lambda$, it is given by:
\begin{equation}
\psi(\varphi) = k d \sin\varphi. 
\tag{5}
\end{equation}

The magnitude of the array factor is thus:
\begin{equation}
|AF(\varphi,\Delta\Phi)|
= 2\left|
\cos\left(
\frac{\pi d}{\lambda}\sin\varphi + \frac{\Delta\Phi}{2}
\right)
\right|.
\tag{6}
\end{equation}

\subsection{Spatial Envelope Obtained by Phase Sweeping}

We now examine the maximum radiated power that can be obtained in each spatial
direction by varying the relative excitation phase $\Delta\Phi$ between the two
radiating elements.
When the relative phase $\Delta\Phi$ is swept over the full interval
$\Delta\Phi\in[0,2\pi]$, the absolute cosine term spans its entire range, and its maximum
value is equal to unity for any $\varphi$. Indeed, the absolute cosine function 
$|\cos(\cdot)|$ is $\pi$-periodic. 
As a consequence, the maximum achievable
array factor magnitude in any spatial direction is:
\begin{equation}
\max_{\Delta\Phi}\left|AF(\varphi,\Delta\Phi)\right| = 2,
\qquad \forall\,\varphi.
\tag{7}
\end{equation}

It follows that the spatial envelope obtained by sweeping the beamforming phase
$\Delta\Phi$ is given by:
\begin{equation}
G_{\mathrm{env}}(\varphi)
=
A_E(\varphi)
+
20\log_{10}(2)
=
A_E(\varphi) + 6~\mathrm{dB}.
\tag{8}
\end{equation}

\begin{figure}[h] 
    \centering
    \includegraphics[width=1\columnwidth]{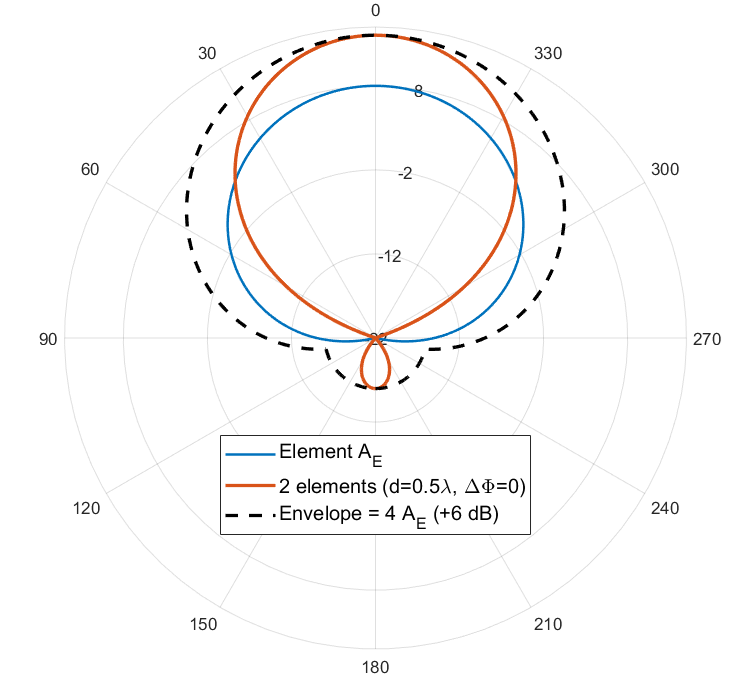} 
    \caption{two-element array pattern envelope}
    \label{model}
\end{figure}
This result shows that, although beamforming allows the main lobe to be steered in
any azimuth direction, the maximum radiated power achievable in each direction is
fully determined by the element radiation pattern. The spatial envelope is therefore
independent of the beamforming phase and is solely governed by the elementary
radiator.

\section{Spatial Upper Bound of Radiated power}
\label{sec:spatial_upper_bound}

We now focus on the spatial distribution of radiated power resulting from the two-element array
model introduced in Section~\ref{sec:two_element_array}. Using the third-order nonlinear signal
model, three spectral regions can be identified, depending on the dominant contribution
to the radiated signal\cite{Rizk2026SpatialUpperBound}.

Independently of the considered region, we are interested in the maximum radiated power
achievable in a given spatial direction when the relative excitation phase $\Delta\Phi$ is swept
over its full range $[0,2\pi]$. This maximum defines a \emph{spatial envelope}, which will be shown
to constitute a deterministic upper bound on the radiated power.
We first consider a single-user transmission scenario,
where the discrete-time complex baseband signal $x(n)$ denotes the useful signal to
be transmitted. The signal is assumed to be zero-mean and identical at the input of
the two RF chains, up to a controllable excitation phase shift.

The output of the power amplifier feeding the first radiating element is modeled by
a memoryless third-order nonlinearity and can be written as:
\begin{equation}
y_1(n) = x(n) + \alpha |x(n)|^2 x(n) + w_1(n),
\tag{9}
\end{equation}
where $\alpha$ denotes the third-order nonlinearity coefficient and $w_1(n)$ is an
additive noise term that accounts for thermal noise and residual impairments.

By convention, the second RF chain applies a relative excitation phase $\Delta\Phi$
to the useful signal. The corresponding PA output is therefore given by:
\begin{equation}
y_2(n) = x(n)e^{j\Delta\Phi}
+ \alpha |x(n)|^2 x(n)e^{j\Delta\Phi}
+ w_2(n),
\tag{10}
\end{equation}
where $w_2(n)$  is assumed to be uncorrelated with
$w_1(n)$.

This signal model explicitly separates three contributions at the antenna inputs:
the useful in-band component proportional to $x(n)$, the third-order intermodulation (IM)
component proportional to $|x(n)|^2x(n)$, and additive noise terms. These three
components will be shown to exhibit distinct spatial behaviors, which motivates the
identification of three corresponding spectral regions in the following analysis.

\subsection{Useful-signal-dominated region}

In the in-band region, the useful signal $x(n)$ dominates. The signals feeding the two radiating
elements are phase-shifted versions of the same waveform and therefore remain fully correlated.

As a result, the radiated fields combine coherently. The directional Equivalent Isotropically Radiated Power (EIRP) can be written as:
\begin{equation}
\mathrm{EIRP}_{\text{sig}}(\varphi,\Delta\Phi)
= P_e+\,A_E(\varphi)
+ 20\log_{10}|AF(\varphi,\Delta\Phi)| 
\tag{11}
\end{equation}
where $P_e$ is the conducted power of an element.
Sweeping the relative phase $\Delta\Phi$ over $[0,2\pi]$ yields the spatial envelope:
\begin{equation}
\mathrm{EIRP}_{\text{sig,env}}(\varphi)
= P_e+\,A_E(\varphi)+ 6~\mathrm{dB}
\tag{12}
\end{equation}
The envelope has exactly the same angular shape as
the single-element radiation pattern, scaled by a factor four.

\subsection{Third-order-distortion-dominated region}

We now consider the adjacent-band region where the third-order
distortion term $\alpha |x(n)|^2 x(n)$ is predominant.
In the present two-element configuration, the input signals applied
to the radiating elements differ only by a deterministic phase shift
$\Delta\Phi$. The third-order distortion term therefore experiences exactly the
same inter-element phase shift as the useful signal. As a consequence,
both components share the same array factor and are beamformed in
the same spatial direction. 
 The corresponding directional EIRP can be expressed as:
\begin{equation}
\mathrm{EIRP}_{\text{IM3}}(\varphi,\Delta\Phi)
= P_{\text{IM3}}+\,A_E(\varphi)
+ 20\log_{10}|AF(\varphi,\Delta\Phi)| 
\tag{13}
\end{equation}

where $P_{\text{IM3}}$ denotes the power of one element associated with the third-order distortion component.

By sweeping $\Delta\Phi$ over $[0,2\pi]$, the spatial envelope becomes:
\begin{equation}
\mathrm{EIRP}_{\text{IM3,env}}(\varphi)
= P_{\text{IM3}}+\,A_E(\varphi)+ 6~\mathrm{dB}
\tag{14}
\end{equation}

Hence, even in the distortion-dominated region, the maximum radiated power is achieved in the
boresight direction and the spatial envelope is again given by the element radiation pattern scaled
by a constant factor. 

\subsection{Noise-dominated region}

Finally, we consider frequency regions where the additive noise terms dominate. In this case, the
noise contributions $w_1(n)$ and $w_2(n)$ feeding the two elements are assumed independent and
uncorrelated. As a consequence, the radiated powers add incoherently. The resulting directional EIRP is given by:
\begin{equation}
\mathrm{EIRP}_{\text{noise}}(\varphi)
= P_{\text{noise}}+\,A_E(\varphi)+ 3~\mathrm{dB}
\tag{15}
\end{equation}

where $P_{\text{noise}}$ denotes the noise power per branch and the +3 dB term accounts for incoherent power addition 
from two uncorrelated noise sources (one per element). This result naturally extends to larger antenna arrays. For an Active Antenna System (AAS) composed of $2MN$ sub-arrays, each generating statistically independent noise contributions, the radiated noise spatial envelope can be expressed as:
\begin{equation*}
\mathrm{EIRP}_{\mathrm{noise,AAS}}(\varphi)
=
P_{\mathrm{noise,s}}
+
A_{\mathrm{sub}}(\varphi)
+
10\log_{10}(2MN)\ .
\end{equation*}
where $P_{\mathrm{noise,s}}$ denotes the noise power per sub-array and $A_{\mathrm{sub}}(\varphi)$ represents the sub-array radiation pattern. Unlike useful signal components and third-order intermodulation (IM3) products, which combine coherently across sub-arrays and therefore scale with a beamforming gain of $20\log_{10}(2MN)$ dB, noise contributions are uncorrelated and add in power rather than in field amplitude. Consequently, their spatial scaling follows a $10\log_{10}(2MN)$ law. This fundamental distinction between coherent field summation and incoherent power addition is central to the spatial upper-bound framework.
Importantly, in contrast to the coherent cases, the spatial distribution of noise does not depend on
the relative phase $\Delta\Phi$. For any value of $\Delta\Phi$, the angular shape of the radiated
noise power remains identical and follows the single-element radiation pattern. The spatial
envelope therefore coincides with this pattern, with a constant $+3$~dB offset due to power
addition.

\subsection{Spatial upper bound}

Considering the three spectral regions together, a key conclusion emerges. Regardless of whether
the radiated power is dominated by the useful signal, by third-order distortion components, or by
noise, the maximum radiated power in any spatial direction is bounded by a deterministic
envelope.

For the two-element array, this envelope has the angular shape of the elementary radiator and
reaches its maximum in the boresight direction. The corresponding maximum level is obtained by
the coherent addition of the two branches when applicable, or by power addition in the noise
regime.

This result establishes the existence of a \emph{spatial upper bound of radiated power}, which is
independent of frequency and of the nature of the radiated signal component. This bound will be
shown in the next section to extend naturally to realistic active antenna systems, where the role of
the elementary radiator is played by the sub-array. This result has been obtained with a simple nonlinearity of order 3, but can be naturally generalized with any order of Volterra model. 

\section{MULTI-USER TRANSMISSION CASE}

We now extend the previous analysis to a multi-user transmission scenario. For clarity, we restrict the discussion to the case of two simultaneously served users, which is sufficient to highlight the key spatial dispersion mechanisms. The results are expected to generalize to $K > 2$ users, but a rigorous proof is left for future work. However, the fundamental mechanisms, power sharing and spatial dispersion, remain valid.

\subsection{Signal model}

Let $u_1(n)$ and $u_2(n)$ denote the discrete-time complex baseband signals associated with user 1 and user 2, respectively. These signals are assumed to be zero-mean and mutually uncorrelated.
At the input of the first radiating element, the composite transmit signal is given by the superposition:
\begin{equation}
x_1(n) = u_1(n) + u_2(n).
\tag{16}
\end{equation}

This signal is fed to a power amplifier modeled by a third-order memoryless nonlinearity. The corresponding PA output can be written as:
\begin{equation}
y_1(n) = x_1(n) + \alpha|x_1(n)|^2x_1(n) + w_1(n),
\tag{17}
\end{equation}
where $\alpha$ denotes the third-order nonlinearity coefficient and $w_1(n)$ is an additive noise term.

For the second radiating element, user-dependent phase shifts are applied in order to form two independent beams. The input signal to the second branch is therefore:
\begin{equation}
x_2(n) = u_1(n)e^{j\Delta\Phi_1} + u_2(n)e^{j\Delta\Phi_2},
\tag{18}
\end{equation}
where $\Delta\Phi_1$ and $\Delta\Phi_2$ denote the beamforming phase shifts associated with user 1 and user 2, respectively.

After the power amplifier, the output signal of the second branch becomes:
\begin{equation}
y_2(n) = x_2(n) + \alpha|x_2(n)|^2x_2(n) + w_2(n),
\tag{19}
\end{equation}
where $w_2(n)$ denotes an additive noise term independent of $w_1(n)$.
Substituting the beamformed inputs into the PA model, the output signals
of the two branches can be written explicitly as
For branch~1,
\begin{equation}
\begin{aligned}
y_1(n) &= u_1(n) + u_2(n) \\
&\quad + \alpha \Big(
|u_1(n)|^2 u_1(n)
+ |u_2(n)|^2 u_2(n) \\
&\qquad\quad
+ |u_1(n)|^2 u_2(n)
+ |u_2(n)|^2 u_1(n) \\
&\qquad\quad
+ u_1^2(n) u_2^*(n)
+ u_2^2(n) u_1^*(n)
\Big)
 + w_1(n)
\end{aligned}
\tag{20}
\end{equation}

For branch~2,
\begin{equation}
\begin{aligned}
y_2(n)
&=
u_1(n)e^{j\Delta\Phi_1}
+u_2(n)e^{j\Delta\Phi_2}
\\
&\quad
+\alpha\Big(
\Gamma_{\mathrm{SD}}(n)
+\Gamma_{\mathrm{XA}}(n)
+\Gamma_{\mathrm{XB}}(n)
\Big)
+w_2(n).
\end{aligned}
\tag{21}
\end{equation}


where $\Gamma_{\mathrm{SD}}(n)$  is the self-distortion:
\begin{equation}
\Gamma_{\mathrm{SD}}(n)
=
|u_1(n)|^2 u_1(n)e^{j\Delta\Phi_1}
+
|u_2(n)|^2 u_2(n)e^{j\Delta\Phi_2}
\tag{22}
\end{equation}

$\Gamma_{\mathrm{XA}}(n)$ is Type-A cross-IM:
\begin{equation}
\Gamma_{\mathrm{XA}}(n)
=
|u_2(n)|^2 u_1(n)e^{j\Delta\Phi_1}
+
|u_1(n)|^2 u_2(n)e^{j\Delta\Phi_2}
\tag{23}
\end{equation}

and $\Gamma_{\mathrm{XB}}(n)$ Type-B cross-IM:
\begin{equation}
\begin{aligned}
\Gamma_{\mathrm{XB}}(n)
&=
u_1^2(n)u_2^*(n)e^{j(2\Delta\Phi_1-\Delta\Phi_2)}
\\
&\quad
+u_2^2(n)u_1^*(n)e^{j(2\Delta\Phi_2-\Delta\Phi_1)}
\end{aligned}
\tag{24}
\end{equation}

Hence, overall IM components contain:
\begin{itemize}
\item Self-distortion terms: proportional to $|u_1(n)|^2u_1(n)$ and $|u_2(n)|^2u_2(n)$;
\item Cross-IM terms of type A: proportional to $|u_1(n)|^2u_2(n)$ and $|u_2(n)|^2u_1(n)$;
\item Cross-IM terms of type B: proportional to $u_1(n)^2u_2^*(n)$ and $u_2(n)^2u_1^*(n)$.
\end{itemize}

The useful signal components $u_1(n)$ and $u_2(n)$ are beamformed toward two distinct spatial directions determined by $\Delta\Phi_1$ and $\Delta\Phi_2$, respectively.

Crucially, the self-distortion terms $|u_k|^2u_k$ preserve the phase relationship of their parent signal $u_k$ and are therefore beamformed in the same direction as the corresponding user.
Similarly, the cross-IM terms of type A also preserve the phase of one of the user signals:
\begin{itemize}
\item The term $|u_2|^2u_1$ carries the phase of $u_1$ and is therefore beamformed in the same direction as user 1;
\item The term $|u_1|^2u_2$ carries the phase of $u_2$ and is therefore beamformed in the same direction as user 2.
\end{itemize}

In contrast, the cross-IM terms of type B involve quadratic phase combinations and do not align with either user beam:
\begin{itemize}
\item The term $u_1^2u_2^*$ at the second element becomes $u_1^2e^{j2\Delta\Phi_1}u_2^*e^{-j\Delta\Phi_2}$, radiating in a direction determined by the phase combination $2\Delta\Phi_1 - \Delta\Phi_2$;
\item The term $u_2^2u_1^*$ radiates in a direction determined by $2\Delta\Phi_2 - \Delta\Phi_1$.
\end{itemize}

The total radiated distortion power is therefore dispersed over four distinct spatial directions:

\begin{enumerate}
\item User 1 beam direction ($\Delta\Phi_1$): useful signal $u_1$ + self-distortion $|u_1|^2u_1$ + cross-IM type A $|u_2|^2u_1$
\item User 2 beam direction ($\Delta\Phi_2$): useful signal $u_2$ + self-distortion $|u_2|^2u_2$ + cross-IM type A $|u_1|^2u_2$
\item Cross-IM direction 1 ($2\Delta\Phi_1 - \Delta\Phi_2$): cross-IM type B $u_1^2u_2^*$
\item Cross-IM direction 2 ($2\Delta\Phi_2 - \Delta\Phi_1$): cross-IM type B $u_2^2u_1^*$
\end{enumerate}

From a spatial-power perspective, the multi-user case is intrinsically more favorable than the single-user case for two fundamental reasons:

\textbf{1. Power sharing:} The total transmit power is divided among multiple beams. Consequently, the peak EIRP in any given direction is reduced compared to the single-user boresight configuration, where all power is concentrated in a single beam.

\textbf{2. Spatial dispersion of distortion:} As demonstrated above, third-order IM products are dispersed over four distinct spatial directions. This dispersion further reduces the radiated power density in any particular direction compared to the single-user case, where all distortion is concentrated in the boresight direction.

For any given direction $\varphi$, the EIRP in the multi-user case therefore satisfies:
\begin{equation}
\text{EIRP}_{\text{MU}}(\varphi) \leq \text{EIRP}_{\text{SU,max}}(\varphi),
\label{eq:mu_bound}
\tag{25}
\end{equation}
where $\text{EIRP}_{\text{SU,max}}(\varphi)$ denotes the maximum EIRP obtained in the single-user boresight case.

This inequality holds for all the components (useful signal, harmonics and noise). Consequently, the spatial upper bound derived for the single-user case in Section III remains \textbf{valid and conservative} in a multi-user transmission scenario. The worst-case radiated power is still achieved in the single-user boresight configuration, and multi-user operation can only reduce the maximum power radiated in any given direction.

\section{Extension to Active Antenna Systems and Experimental Validation}
\label{sec:extension_aas_measurements}

In this section, we extend the spatial upper-bound concept derived for the two-element array to a
realistic AAS. We then validate the theoretical findings using
OTA measurements performed on an antenna previously
used in our earlier work\cite{Rizk2026SpatialUpperBound}.
The AAS under test is the SPEAR radio unit from Andrew, and is stimulated by OpenAirInterface (OAI) software\cite{Kaltenberger2025OAI6G}.
\subsection{Extension of the spatial upper bound to AAS}
In a practical AAS, radiating elements are grouped into 
sub-arrays, each driven by an independent RF chain. Let $M$ denote the number 
of sub-array rows and $N$ the number of columns. Assuming two orthogonal 
polarizations ($\pm 45^\circ$), the total number of RF chains is equal to 
$2MN$.

The EIRP in a given
direction $(\theta,\varphi)$ is given by \cite{3GPP38922}:

\begin{equation}
\begin{aligned}
\mathrm{EIRP}_{\text{sig}}(\theta,\varphi,\Delta\Phi)
&= P_{\mathrm{sub}}+3~\mathrm{dB}+ A_{\mathrm{sub}}(\theta,\varphi) \\
&\quad + 20\log_{10}\!\left|AF_A(\theta,\varphi,\Delta\Phi)\right| 
\end{aligned}
\tag{26}
\end{equation}

where $P_{\mathrm{sub}}$ is the conducted power of one sub-array, +3 dB accounts for the 2 polarizations,
$A_{\mathrm{sub}}(\theta,\varphi)$ is the sub-array radiation pattern, and
the array factor is defined as:

\begin{equation}
AF_{\mathrm{A}}(\theta,\varphi)
=
\sum_{m=0}^{M-1}
\sum_{n=0}^{N-1}
w_{m,n}\,
v_{m,n}(\theta,\varphi)
\label{eq:AF}
\tag{27}
\end{equation}

with
\[v_{m,n}(\theta,\varphi)=\exp\!\left(
j2\pi\left[m\frac{d_v}{\lambda}\cos\theta
+
n\frac{d_h}{\lambda}\sin\theta\,\sin\varphi\right]
\right)\]

and
\[w_{m,n}=\exp\!\left(j\left[m\Delta\Phi_V
+n\Delta\Phi_H\right]\right),\]
in which $\Delta\Phi_V$ and $\Delta\Phi_H$ are respectively the relative excitation phase in vertical and horizontal direction.
The spatial upper-bound concept derived for the two-element array naturally extends to an AAS. Indeed, for a given observation direction $(\theta_0,\varphi_0)$, the geometrical phase term is fully deterministic.
Since the excitation phase gradients $\Delta\Phi_V$ and $\Delta\Phi_H$ are controllable parameters, they can always be selected to compensate the geometrical phase progression.
Specifically, choosing:
\begin{equation*}
\Delta\Phi_V = -2\pi \frac{d_v}{\lambda}\cos\theta_0 ,
\quad
\Delta\Phi_H = -2\pi \frac{d_h}{\lambda}\sin\theta_0\sin\varphi_0
\end{equation*}
yields:
\begin{equation*}
m\Delta\Phi_V + n\Delta\Phi_H
=
-2\pi
\left[
m\frac{d_v}{\lambda}\cos\theta_0
+
n\frac{d_h}{\lambda}\sin\theta_0\sin\varphi_0
\right],
\end{equation*}
which exactly cancels the geometrical phase term in $v_{m,n}(\theta_0,\varphi_0)$.
Consequently, all phasors in the double summation become co-phased and add constructively.
The array factor then reduces to:
\begin{equation*}
AF_A(\theta_0,\varphi_0)=MN.
\end{equation*}
Following the reasoning developed in Sections~\ref{sec:two_element_array} and
\ref{sec:spatial_upper_bound}, the elementary radiator in the spatial upper-bound analysis is no
longer a single element, but the sub-array itself. Consequently, the spatial envelope governing the
maximum radiated power over all beamforming configurations has the angular shape of the
sub-array radiation pattern.

\subsection{Measurement setup and frequency-domain characterization}

The experimental validation is based on OTA measurements performed in an anechoic chamber
using a realistic Massive MIMO antenna operating in the 3.4--3.8\,GHz band\cite{Rizk2026SpatialUpperBound}. The antenna is first
configured in a single-user transmission mode with a fixed boresight beam.
The received signal is measured in an azimuth cut for $ \varphi \in [-60^\circ, 60^\circ]$, using a calibrated probe
antenna. The received power spectral density is computed by integrating the received signal power
over a bandwidth of 1\,MHz. This operation is repeated over the frequency range from 3.4\,GHz
to 4\,GHz, resulting in a spatial and frequency-dependent representation of the Power Spectral
Density (PSD).
\begin{figure}[h]
  \centering
  \includegraphics[width=1\linewidth]{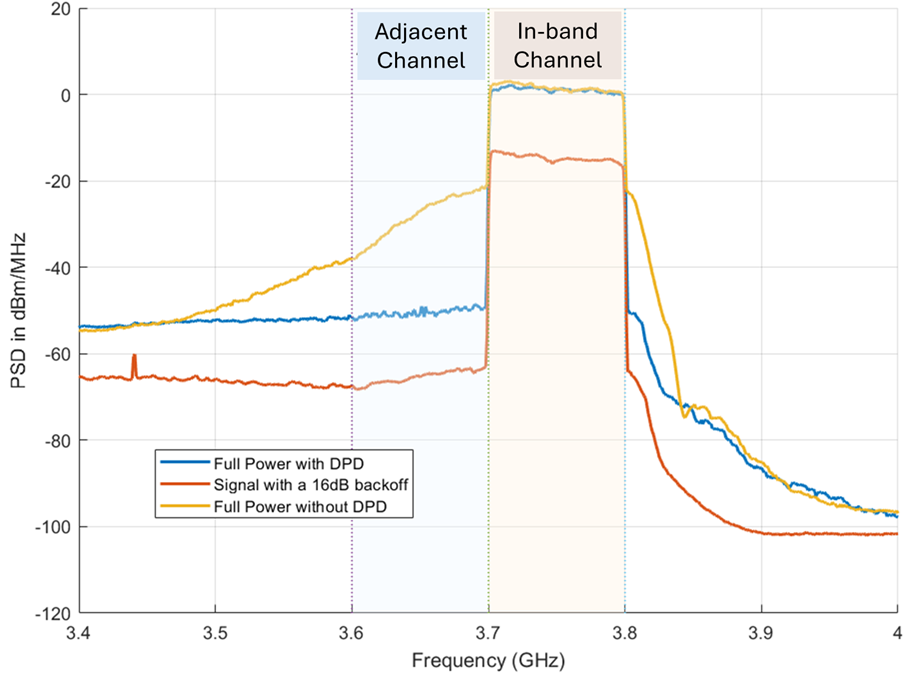}
  \caption{Measured boresight EIRP spectral density (1 MHz RBW) from 3.4 to 4.0 GHz for three configurations (Full power with and without DPD, 16 dB backoff). Vertical dashed lines indicate measurement frequencies for angular analysis: 
3.75 GHz (in-band region), 3.65 GHz (IM3 and noise dominated region).}
  \label{fig:boresight_psd}
\end{figure}
Figure 3 shows the measured power spectral density in boresight for 3 signal configurations:
\begin{itemize}
\item full power 5G NR 100 MHz signal centered at 3.75 GHz with DPD
\item  full power 5G NR 100 MHz signal without DPD
\item 5G NR 100 MHz signal with a 16 dB backoff.
\end{itemize}
Full power transmission corresponds to a TX power of 72 dBm per polarization (75 dBm in total). In this experiment, the probe is used with only one polarization, and the attenuation between the AAS and the probe is 51 dB. This results in a received PSD of approximately 1 dBm/MHz in the band (from 3.7 to 3.8 GHz). 
These curves provide in each configuration a direct experimental estimate of the maximum
radiated power as a function of frequency and serve as a reference upper bound for the angular
analysis that follows.
\begin{figure}[h]
  \centering
  \includegraphics[width=1\linewidth]{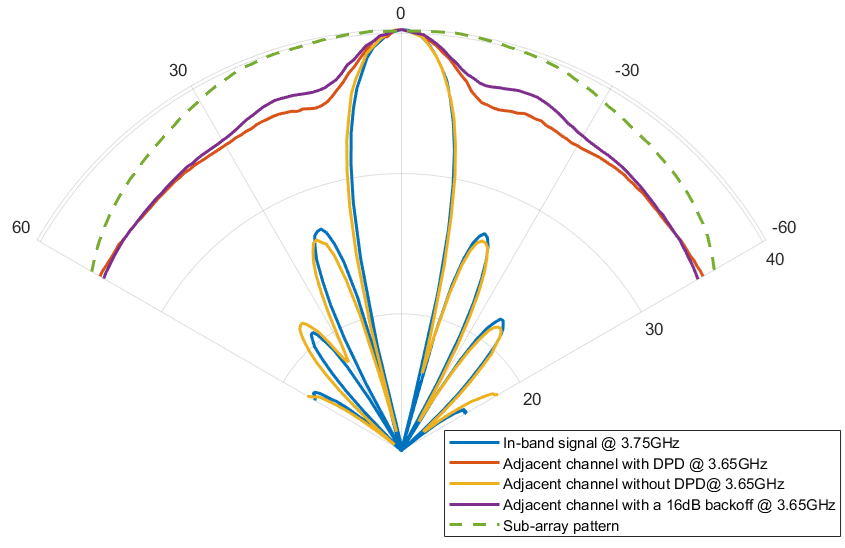}
  \caption{Azimuth cuts for boresight beam, signals integrated on 100 MHz (in-band, adjacent with and without DPD, adjacent with a 16 dB backoff)}
  \label{fig:Azimut cuts}
\end{figure}

Figure 4 shows the spatial dispersion (azimuth cuts in fixed boresight beam configuration) for: \begin{itemize}
\item the full-power in-band signal
\item  Adjacent channel of the signal without DPD: the shape is quite similar as for the in-band signal, confirming that the IM signals are beamformed in the same direction as the in-band signal
\item Adjacent channel of the signal with DPD: the DPD reduces dramatically the IM signals, and allows a spatial dispersion of the adjacent band signal
\item  Adjacent channel of the signal with a 16 dB backoff: in this case the level of IM is quite low, providing also spatial dispersion
\item sub-array pattern, which corresponds to the spatial bound.
\end{itemize}
All radiation patterns are normalized to  boresight level (0 dB reference), allowing direct visual comparison of
the spatial distribution between useful signal and out-of-band
emissions. This normalization highlights the relative spatial
dispersion mechanisms, independently
of absolute power levels. 

In a second experiment, radiation patterns are measured for various beam 
directions ($-42^\circ$, $-30^\circ$, $-18^\circ$ $0^\circ$, $18^\circ$, $30^\circ$ and $60^\circ$), using full power signal with DPD.
Figure~5 shows the radiation patterns (azimuth cuts) of the in-band and adjacent  spectral components for various beam directions, together with the theoretical spatial upper bound derived from the 
boresight level, and the shape of a theoretical sub-array which follows the 3GPP antenna modeling
framework
\cite{3GPP38803} with $\varphi_{3\mathrm{dB}}=85^\circ$.

\begin{figure}[h]
  \centering
  \includegraphics[width=1\linewidth]{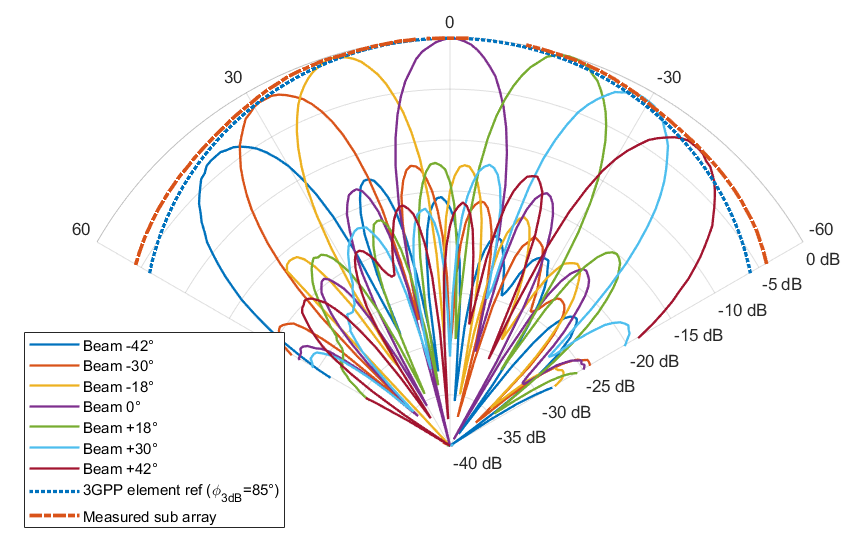}
 \includegraphics[width=1\linewidth]
 {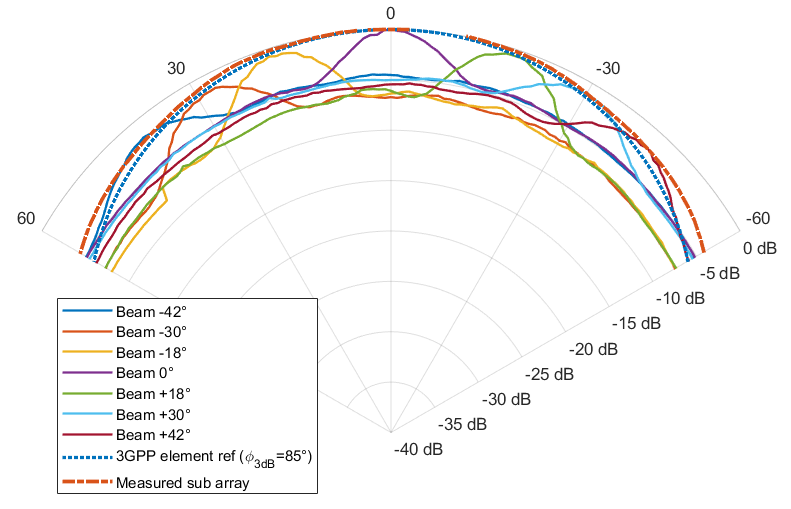}
  \caption{Validation of spatial upper bound for seven beam steering angles (-42°, -30°, -18°, 0°, 18°, 30°, 60°). 
The maximum deviation between the bound and measurements is below 1~dB.}
  \label{fig:seven beams}
\end{figure}

An excellent agreement is observed between the theoretical 
bound and the measured radiation levels across all beam 
steering configurations. Quantitatively, the maximum deviation 
between the theoretical spatial envelope and measured EIRP 
is less than 1 dB for all seven beam directions in both 
spectral regions (in-band and adjacent band). This remarkable 
agreement validates three key aspects of the proposed framework:
1) The spatial upper bound derived from boresight measurement 
   accurately predicts the maximum EIRP envelope for arbitrary 
   beam steering directions;
2) The 3GPP sub-array radiation pattern model ($\varphi_{3\mathrm{dB}}=85^\circ, 
   A_m=$30 dB) provides an accurate representation of the 
   elementary radiator characteristics of our antenna;
3) The theoretical prediction that maximum radiated emission 
   always occurs at boresight holds regardless of beam 
   steering angle.
Concerning the adjacent band, the signal (with DPD) is a mixture of 
residual IM components and noise. The sub-1 dB margin 
between theory and measurement is well within the expected 
OTA uncertainty (±1.3 dB for in-band, ±3.0 dB for out-of-band 
per 3GPP TS 38.141-2), confirming the conservativeness and 
practical applicability of the proposed spatial upper bound 
framework.

\subsection{Experimental Validation in the Multi-User MIMO Case}

The multi-user (MU) scenario was experimentally evaluated using the antenna configured to simultaneously serve two users steered toward distinct azimuth directions. The users were pointed toward $\varphi_1 = 0^\circ$ and $\varphi_2 = 18^\circ$.

For each azimuth angle, the radiated power was computed by integrating the PSD over the in-band and adjacent band.
Measurements were performed with and without DPD in order to clearly observe the spatial components of the adjacent band.
According to the model used in Section~IV, the nonlinear output of each branch contains:

\begin{enumerate}
    \item Self-distortion terms and Type A cross-IM, proportional to $|u_k|^2 u_k$ and $|u_\ell|^2 u_k$, which preserve the phase gradient of one of the users and therefore radiate in the original user directions $\varphi_1$ and $\varphi_2$.
    
    \item Type B cross-IM terms, proportional to $u_1^2 u_2^*$ and $u_2^2 u_1^*$, which generate additional phase gradients:
    \begin{equation}
    \Delta \Phi_{B1} = 2\Delta \Phi_1 - \Delta \Phi_2,
    \qquad
    \Delta \Phi_{B2} = 2\Delta \Phi_2 - \Delta \Phi_1.
    \tag{28}
    \end{equation}
\end{enumerate}

For an AAS steered in azimuth, where the phase gradient satisfies $\Delta \Phi(\varphi) \propto \sin\varphi$, the corresponding directions are:
\begin{equation}
\sin \varphi_{B1} = 2 \sin \varphi_1 - \sin \varphi_2,
\qquad
\sin \varphi_{B2} = 2 \sin \varphi_2 - \sin \varphi_1.
\tag{29}
\end{equation}

For $\varphi_1 = 0^\circ$ and $\varphi_2 = 18^\circ$, this yields:
\begin{equation}
\varphi_{B1} \approx -18^\circ,
\qquad
\varphi_{B2} \approx 38.2^\circ.
\tag{30}
\end{equation}
Therefore, in addition to the two useful beams, two additional spatial directions are theoretically expected for third-order IM products.
\begin{figure}[h]
  \centering
  \includegraphics[width=1\linewidth]{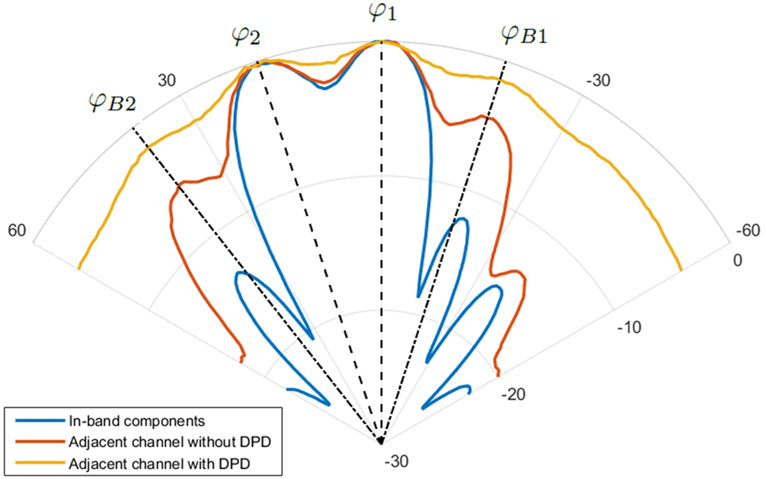}
  \caption{Multi-user spatial dispersion. All radiation patterns (in-band and adjacent band) are normalized to the in-band boresight level (0 dB reference), enabling direct comparison of spatial distribution shapes.}
  \label{fig:MU dispersion}
\end{figure}
Figure 6 shows the spatial dispersion of both in-band and adjacent-band radiation for the multi-user configuration. All radiation patterns are normalized to the boresight level (0 dB reference), allowing direct visual comparison of 
the spatial distribution between useful signal and out-of-band emissions. This normalization highlights the relative spatial dispersion mechanisms described in Section IV, independently 
of absolute power levels.
The in-band radiation exhibits two main lobes aligned with the user directions ($0^\circ$ and $18^\circ$), as expected from the beamforming configuration.
The adjacent band radiation without DPD shows a clear angular structure with maxima partially aligned with the user beams, consistent with the Type~A distortion terms described above. Evidence of additional spatial spreading is also observed, consistent with the predicted Type~B mechanisms. This spatial distribution is mitigated with the DPD, which reduces the level of IM.

These experimental observations validate the analytical framework developed in Section~IV. In particular, they confirm that:
\begin{itemize}
    \item third-order nonlinear products in MU transmission generate additional spatial components,
    \item the spatial distribution of OOB radiation is inherently more dispersed in MU compared to the single-user case.
\end{itemize}

This work does not aim at defining a normative 3GPP-compliant limit. 
Instead, it proposes a conservative engineering framework to derive 
a spatial upper bound from boresight measurements for coexistence analyses.

The bound accounts for three uncertainty sources:
(i) OTA measurement uncertainty,
(ii) sub-array radiation pattern measurement uncertainty, and
(iii) inter-sub-array variability.

According to 3GPP TS~38.141-2 (Release~18), the maximum acceptable OTA 
uncertainty for FR1 radiated transmit power measurements in the range 
$3~\mathrm{GHz} < f \leq 7.125~\mathrm{GHz}$ is $\pm 1.3$~dB (95\% confidence). 
A conservative one-sided bound is therefore defined as
\begin{equation}
\mathrm{EIRP}_{\mathrm{bound}}(f)
=
\mathrm{EIRP}_{\mathrm{meas}}(f)
+
1.3~\mathrm{dB}.
\tag{31}
\end{equation}
For out-of-band and spurious quantities, absolute-power uncertainties 
may reach up to $3$~dB. When a uniform conservative treatment over the 
entire spectrum is required, a $3$~dB margin may be applied.

Sub-array pattern measurements are subject to the same OTA uncertainty. 
In addition, manufacturing dispersion between sub-arrays introduces 
angular-dependent deviations. These effects can be addressed either 
through worst-case characterization or by introducing an additional 
statistically derived margin.

The resulting bound shall therefore be interpreted as a conservative 
spatial envelope derived from the measured boresight EIRP and the 
elementary radiator characteristics, augmented by explicitly stated 
uncertainty margins.

\section{Discussion and Implications for Coexistence}

The proposed spatial upper bound has three key implications for
coexistence assessments involving AAS.

\subsection{Independence from correlation modeling}

The spatial upper bound does not require explicit knowledge of the
spatial correlation between RF chains. It is fully characterized by:
(i) the maximum power spectral density measured in the boresight
direction, and (ii) a deterministic spatial envelope governed by the
elementary radiator (element or sub-array).

Correlation effects, often quantified by a factor $\rho$\cite{3GPP38922}, mainly affect
the degree to which unwanted emissions are beamformed.
However, this influence is implicitly captured in the measured
boresight spectrum. When nonlinear distortion increases, the adjacent-channel leakage ratio
(ACLR) degrades, and any degradation due to correlated distortion components is reflected uniformly in the spatial envelope, which
bounds the radiated emissions in all directions. As a consequence, the upper bound
remains valid and conservative without requiring explicit modeling
of distortion correlation mechanisms.

\subsection{Role of DPD}

DPD plays a dual role. First, it reduces
 unwanted signals in conducted mode by improving ACLR. Second,
by suppressing correlated distortion components, it limits their
coherent beamforming in the radiated domain.
When distortion becomes noise-like, emissions add predominantly
incoherently, resulting in a flatter spatial distribution. Hence,
DPD reduces both the absolute emission level and its spatial
concentration. This dual effect highlights the central role of DPD in
modern Massive MIMO transmitters, not only for in-band linearity but also for spatial control of out-of-band emissions.

\subsection{Practical impact for coexistence studies}

From a regulatory and system-engineering perspective, the main
benefit of the proposed framework is methodological simplification.
Rather than modeling detailed beamforming states or nonlinear
distortion mechanisms, a conservative upper bound can be derived
directly from boresight measurements at maximum transmit power.

This measurement-oriented approach preserves physical rigor while
substantially reducing the complexity of interference assessments.

\section{Conclusion}
\label{sec:conclusion}

This paper established the existence of a deterministic 
spatial upper bound for the radiated power of Active Antenna 
Systems (AAS) employing beamforming and Massive MIMO.

Starting from a two-element array with third-order 
nonlinearities, we showed that sweeping the excitation phase 
defines a spatial envelope whose angular shape coincides with 
that of the elementary radiator. The result naturally extends 
to realistic AAS architectures, where the sub-array acts as 
the elementary radiating building block. In all cases, the 
maximum radiated emission is achieved in the boresight 
direction.

A central outcome of this work is that the proposed bound 
is independent of frequency, signal nature (useful signal, 
nonlinear distortion, or noise), beamforming configuration, 
and explicit correlation modeling between RF chains. All 
relevant effects are implicitly captured by the measured 
boresight spectral density.

Experimental validation on a 3.5 GHz Massive MIMO antenna 
confirmed the theoretical predictions across three spectral 
regions and seven beam steering angles, with margins 
consistent with OTA measurement uncertainty. Multi-user 
transmission was shown to inherently reduce peak EIRP through 
power sharing and spatial dispersion of intermodulation 
products, validating the predicted four-direction radiation 
pattern.

The resulting framework provides a conservative and 
experimentally accessible methodology for coexistence 
assessments involving beamformed radio systems, particularly 
relevant for adjacent-band scenarios such as 5G/radio 
altimeter interference mitigation.

\nocite{*}
\balance
\bibliographystyle{IEEEtran}
\bibliography{ref1}

@article{Larsson2018OOBClarified,
  author  = {Larsson, Erik G. and {Van der Perre}, Liesbet},
  title   = {{Out-of-Band Radiation From Antenna Arrays Clarified}},
  journal = {IEEE Wireless Communications Letters},
  volume  = {7},
  number  = {4},
  pages   = {610--613},
  month   = aug,
  year    = {2018},
  doi     = {10.1109/LWC.2018.2802519}
}

@techreport{Rizk2026SpatialUpperBound,
  author      = {Rizk, Christ and Nussbaum, Dominique and Seguenot, Eric and Kaltenberger, Florian and Moro, Andrea and Sinicco, Alessandro},
  title       = {{Actual Out-of-Band Emissions from Massive MIMO Antennas}},
  institution = {EURECOM},
  address     = {Sophia Antipolis, France},
  year        = {2025},
  month       = oct,
  note        = {Presented Feb. 2026},
  url         = {https://www.eurecom.fr/en/publication/8396}
}

@book{balanis2016,
  author    = {Balanis, Constantine A.},
  title     = {{Antenna Theory: Analysis and Design}},
  edition   = {4th},
  publisher = {Wiley},
  address   = {Hoboken, NJ, USA},
  year      = {2016},
  isbn      = {978-1-118-64206-1}
}

@techreport{3GPP38803,
  author      = {{3GPP}},
  title       = {{Study on New Radio (NR) Access Technology -- RF and Coexistence Aspects}},
  institution = {3rd Generation Partnership Project},
  type        = {Technical Report},
  number      = {TR 38.803},
  year        = {2024},
  note        = {Release 14},
  url         = {https://www.3gpp.org}
}

@techreport{3GPP38922,
  author      = {{3GPP}},
  title       = {{Study on International Mobile Telecommunications (IMT) Parameters}},
  institution = {3rd Generation Partnership Project},
  type        = {Technical Report},
  number      = {TR 38.922},
  year        = {2025},
  note        = {Release 19},
  url         = {https://www.3gpp.org}
}

@book{Hansen2009PhasedArray,
  author    = {Hansen, Robert C.},
  title     = {{Phased Array Antennas}},
  publisher = {Wiley},
  year      = {2009}
}

@article{Kaltenberger2025OAI6G,
  author  = {Kaltenberger, Florian and Melodia, Tommaso and Ghauri, Ijaz and Polese, Michele and Knopp, Raymond and Nguyen, Thanh Tu and Velumani, S. and Villa, Davide and Bonati, Luca and Schmidt, Robert and Arora, Sagar and Irazabal, Maria and Nikaein, Navid},
  title   = {{Driving Innovation in 6G Wireless Technologies: The OpenAirInterface Approach}},
  journal = {Computer Networks},
  volume  = {269},
  pages   = {111410},
  month   = sep,
  year    = {2025},
  doi     = {10.1016/j.comnet.2025.111410}
}
\vspace{12pt}
\end{document}